\documentclass[aps,twocolumn,preprintnumbers,showpacs,showkeys,nofootinbib,superscriptaddress]{revtex4}
\usepackage{amsmath} \usepackage{graphicx} \usepackage{amsfonts}
\usepackage{array} \usepackage{amsthm} \usepackage{bm}
\usepackage{palatino} \usepackage{mathpazo} 
\usepackage{supertabular}

\usepackage[breaklinks]{hyperref}
\usepackage{color}

\usepackage{epstopdf}

\newcommand{\nn}{\nonumber}
\newcommand{\be}{\begin{equation}}
\newcommand{\ee}{\end{equation}}
\newcommand{\ba}{\begin{eqnarray}}
\newcommand{\ea}{\end{eqnarray}}
\newcommand{\bal}{\begin{align}}
\newcommand{\eal}{\end{align}}

\newcommand{\al}{\alpha}

\newcommand{\La}{\Lambda}
\newcommand{\bt}{\beta}

\newcommand{\ga}{\gamma}

\newcommand{\ta}{\theta}

\newcommand{\de}{\delta}

\newcommand{\bw}{\begin{widetext}}
\newcommand{\ew}{\end{widetext}}

\begin{document}

\title{Astrophysical flows near  $f(T)$ gravity black holes}

\author{Ayyesha K. Ahmed}
\email{ayyesha.kanwal@sns.nust.edu.pk}\affiliation{Department of Mathematics, School of Natural Sciences (SNS), National University of Sciences and Technology (NUST), H-12, Islamabad, Pakistan}

\author{Mustapha Azreg-A\"{\i}nou}\email{azreg@baskent.edu.tr}
\affiliation{Engineering Faculty, Ba\c{s}kent University, Ba\u{g}l\i ca Campus, Ankara, Turkey}

\author{Sebastian Bahamonde}
\email{sebastian.beltran.14@ucl.ac.uk}\affiliation{Department of Mathematics, University College London,
	Gower Street, London, WC1E 6BT, UK}

\author{Salvatore  Capozziello}
\email{capozziello@na.infn.it}
\affiliation{Dipartimento di Fisica, Universit\'a di Napoli
	\textquotedblleft{Federico II}\textquotedblright, Compl. Univ. di Monte S. Angelo, Edificio G, Via
	Cinthia, I-80126,  Napoli, Italy}
\affiliation{Gran Sasso Science Institute (INFN), Via F. Crispi 7, I-67100, L' Aquila,
	Italy}
\affiliation{INFN Sez. di Napoli, Compl. Univ. di Monte S. Angelo, Edificio G, Via
	Cinthia, I-80126,
	Napoli, Italy}

\author{Mubasher  Jamil}\email{mjamil@sns.nust.edu.pk}
\affiliation{Department of Mathematics, School of Natural
	Sciences (SNS), National University of Sciences and Technology
	(NUST), H-12, Islamabad, Pakistan}

\begin{abstract}
In this paper, we study the accretion process for fluids flowing near  a  black hole in the context of $f(T)$ teleparallel gravity. Specifically, by performing a dynamical analysis by a Hamiltonian  system,  we are able to find the sonic points. After that, we consider different isothermal test fluids in order to study the accretion process when they are falling onto the black hole. We found that these flows can be classified according to the equation of state and the black hole features. Results are compared in $f(T)$ and $f(R)$ gravity.

\end{abstract}

\date{\today}


\maketitle

\section{Introduction}
One of the most important problems in modern cosmology  is  to address the dark energy issue, which is  responsible for the accelerated   expansion of the observed Universe. Over the last few decades, several studies have been focused on trying to tackled this problem. It is well known that this form of energy is acting as a repulsive gravitational force so that in General Relativity (GR) one needs to consider a further non-standard fluid with a negative pressure to justify this accelerated scenario. The simplest approach is to consider a cosmological constant in order to explain it. However from quantum considerations, the necessary expected value of it must be extremely  larger than the observed value \cite{Martin:2012bt}. Another approach to  the cosmic accelerated behavior  comes from modified theories of gravities where, instead of searching for new material ingredients, the philosophy is to address cosmic dynamics taking into account possible further degrees of freedom of the gravitational field. A  very well-studied approach to modified  gravity comes out from  the ``Teleparallel equivalent to General Relativity" (TEGR) . This theory yields the same field equations as in General Relavity, so that TEGR is an alternative and equivalent theory to it. However, the geometrical interpretation of these theories are different. On the one hand, GR assumes a non-zero curvature and a vanishing torsion by choosing the symmetric Levi-Civita connection. On the other hand, TEGR considers an antisymmetric connection provided with a non-vanishing torsion and a zero curvature
(Weitzenb\"ock connection). In other words, one can say that GR uses the curvature to geometrize the space-time, meanwhile TEGR uses torsion  to explain  gravitational effects.  In TEGR, we need to use tetrad fields as the dynamical variables in order to define the Weitzenb\"ock connection (See \cite{Obukhov:2002tm,Arcos:2005ec,Maluf:2013gaa,Weit,Cho,Cho2,Hayashi,Hay,Kop,Hammond:2002rm,Aldrovandi:2013wha}, and also the review \cite{Cai:2015emx} for the basis in TEGR).

A natural generalization of TEGR is, instead of using the scalar torsion $T$, to consider an arbitrary and smooth function of the torsion $f(T)$ in the gravitational action \cite{Ferraro:2006jd,Bengochea:2008gz,Li:2010cg,Sotiriou:2010mv}. This theory is the so-called ``$f(T)$ gravity". The idea comes out naturally exactly as when GR is generalized to  $f(R)$ gravity \cite{Capozziello2002,Capozziello:2011et,Sotiriou:2008rp,Nojiri:2010wj}. An important problem related to $f(T)$ gravity is that it is no longer invariant under local Lorentz transformations so that different tetrads might give rise to different solutions. Therefore one needs to be very careful choosing the correct tetrad \cite{Tamanini:2012hg}. Although TEGR is equivalent to GR, it is important to mention that $f(R)$ is no longer equivalent to $f(T)$ gravity \cite{GW}.  One needs to consider a more general theory of gravity, the so-called ``$f(T,B)$ gravity" to obtain the teleparallel equivalent to $f(R)$ gravity \cite{Bahamonde:2015zma}. In addition, it is important to remark that $f(T)$ gravity contains only second order derivative terms meanwhile $f(R)$ gravity contains up to fourth order derivative terms in the metric formalism.

In the last few years, $f(T)$ gravity acquired a lot of interest in cosmology due to the possibility to explain by it the accelerated expansion of the cosmic Hubble fluid (see \cite{Myrzakulov:2010vz,Chen:2010va,Bamba:2010wb,Wu:2010mn,Farajollahi:2011af,Bengochea:2010sg,Jamil:2012nma,Jamil:2012vb,Jamil:2012fs,Bamba:2012rv}). In addition, astrophysical studies related with compact objects as black holes has been considered among $f(T)$ gravity such as in \cite{Boehmer:2011gw,Ferraro:2011ks,HamaniDaouda:2011iy,Wang:2011xf,Paliathanasis:2014iva}. {However, it is worth noticing that this   is not the only solution that can be achieved by the Noether Symmetry Approach. As shown in \cite{felice} for $f(R)$ gravity, the symmetries select the form of the function and several Noether vectors can exist. In the specific case of $f(T)$ gravity, other solutions have been found as discussed  in \cite{1,2}}. A very studied  process, known as accretion,  occurs when a fluid is situated in the vicinity of a black hole or a massive astrophysical object (see \cite{1u,2u,W1,W2}). In this process, the compact object takes particles from the fluid and  increases its mass. Accretion takes place regularly in the Universe, and it can be used to test gravitational theories using observational measurements \cite{Harko:2009rp,Pun:2008ae,Perez:2012bx}. The first study on accretion was performed using Newtonian gravity by Bondi  \cite{4u}. He found transonic solutions for a gas accreting onto compact objects. Michel extended the later work considering GR for a Schwarzschild black hole \cite{5u}. An important work in this field has been pursued by Babichev et. al, where they showed that the mass of the black hole decreases when a phantom fluid is in accretion onto it \cite{Babichev:2004yx}. Later, M.~Jamil and A.~Qadir showed that primordial black holes  decay earlier when the effect of accretion of phantom energy is considered \cite{Jamil:2009qs}. In addition, B.~Nayak and M.~Jamil also found that primordial black holes  accrete radiation, matter and vacuum energy when they pass through radiation, matter and vacuum dominated eras, respectively, with the result  that they live longer during the ration era \cite{Nayak:2011sk}. After that, several works have been done on  accretion onto compact objects (see \cite{Das:2002ce,Chakrabarti:1997ht,Debnath:2015hea,Bahamonde:2015uwa,Babar:2015kaa}).

Recently, A. K. Ahmed and collaborators studied accretion for cyclic and heteroclinic flows near  $f(R)$ black holes \cite{Ahmed:2015tyi}. In this paper, we will use a similar formalism in order to study the accretion process in a black hole in the context of  $f(T)$ gravity.

This paper is organized as follows: In Section II, we briefly introduce the TEGR and $f(T)$ gravity. In Section III, we discuss  the metric representation of black holes in $f(T)$ gravity.  Section IV is devoted to find the general equations for spherical accretion. In Section V, we perform a dynamical system analysis using the Hamiltonian formalism and we study the system at the critical points (CPs). In Section VI, we obtain solutions for isothermal test fluids for different kind of fluids. In Section VII, we analyze the accretion process for a polytropic test fluid. Finally, in Section VIII, we discuss our results and draw conclusions. Throughout the paper we will  use the  metric signature  $(-,+,+,+)$ and the geometric units $G=c=1$.

\section{Teleparallel equivalent of General Relativity and $f(T)$ gravity}
Let us briefly introduce TEGR and its generalization which is the so called $f(T)$ gravity. We will adopt the notation used in \cite{Bahamonde:2015zma}. In this theory, the dynamical variable is the tetrad field $e_{a}^{\mu}$ (or vierbein), where Latin and Greek index indicate  tangents space  and space-time index respectively. The construction of this theory lies on the relationship between the tetrad field and the metric $g_{\mu\nu}$ in the following way
\begin{align}
 g_{\mu\nu} &= e^{a}_{\mu} e^{b}_{\nu} \eta_{ab} \,, \\
 g^{\mu\nu} &= E^{\mu}_{a} E^{\nu}_{b} \eta^{ab} \,,
\end{align}
where $g^{\mu\nu}$ is the inverse of the metric, $E^{\mu}_{a}$ is the inverse tetrad which satisfies the relation $E^{\mu}_{a}e^{a}_{\nu}=\delta_{\nu}^{\mu}$ and $\eta_{ab}=(-1,1,1,1)$ is the Minkowski metric. Therefore, at each point $x^{\mu}$ of the manifold, the tetrad field form an orthonormal basis for the tangent space.\\
As we discussed before, TEGR uses a specific connection (Weitzenb\"ock connection) where the space-time is globally flat but it is endowed with a nonzero torsion tensor. This connection is defined by
\begin{align}
W_{\mu}{}^{\lambda}{}_{\nu}&=E^{\lambda}_{a}\partial_{\mu}e^{a}_{\nu}\,.
\end{align}
Then, we can construct the torsion tensor using the antisymmetric part of the Weitzenb\"ock connection as follows
\begin{align}
T^{\lambda}{}_{\mu\nu} &= W_{\mu}{}^{\lambda}{}_{\nu} - W_{\nu}{}^{\lambda}{}_{\mu} =E_{a}^{\lambda}\Big(
\partial_{\mu} e_{\nu}^{a} - \partial_{\nu}e_{\mu}^{a}\Big) \,.
\end{align}
Using the torsion tensor, one can define the contorsion tensor
\begin{align}
K_{\mu}\,^{\lambda}\,_{\nu}&=\frac{1}{2}\Big(T^{\lambda}\,_{\mu\nu}-T_{\nu\mu}\,^{\lambda}+T_{\mu}\,^{\lambda}\,_{\nu}\Big)\,.
\end{align}
In addition, it is useful to define
\begin{align}
S^{\mu\nu\lambda}=\frac{1}{4}(T^{\mu\nu\lambda}-T^{\nu \mu \lambda }-T^{\lambda \mu \nu})+\frac{1}{2}(g^{\mu\lambda}T^\nu-g^{\mu\nu}T^\lambda)\,,
\end{align}
where $T^{\mu}=T^{\lambda}{}_{\lambda}{}^{\mu}$ is the contraction of the torsion tensor. \\
Using the above tensor, the torsion scalar $T$ can be defined as
\begin{align}
T=S_{\mu}{}^{\nu\lambda}T^{\mu}{}_{\nu\lambda}.
\end{align}
The Riemann tensor can be expressed depending on the contorsion tensor as follows
\begin{align}
R^{\lambda}\,_{\mu\sigma\nu} = \nabla_{\nu}K_{\sigma}{}^{\lambda}{}_{\mu} -
\nabla_{\sigma}K_{\nu}{}^{\lambda}{}_{\mu} +
K_{\sigma}{}^{\rho}{}_{\mu}K_{\nu}{}^{\lambda}{}_{\rho} -
K_{\sigma}{}^{\lambda}{}_{\rho}K_{\nu}{}^{\rho}{}_{\mu} \,.
\end{align}
Here $\nabla_{\mu}$ represents the covariant metric derivative. Therefore, the Ricci scalar $R$ and the torsion scalar $T$ are related by
\begin{align}
R&=-T+\frac{2}{e}\partial_{\mu}(e T^{\mu})\,,\label{RTB}
\end{align}
where $e=\textrm{det}(e^{a}_{\mu})$. It is important to remark that $B=\frac{2}{e}\partial_{\mu}(e T^{\mu})$ is a boundary term. \\
Instead of using the Ricci scalar $R$ as in GR, the TEGR Lagrangian density is described by the torsion scalar $T$
\begin{align}
S_{\rm TEGR} = \int T e\, d^4x \, .
\end{align}
Since $B$ is a boundary term, from \eqref{RTB}, one can see that the TEGR action will arise to the same field equations as the Einstein-Hilbert action, making these two theories equivalent. \\
One important and very well-studied generalization of TEGR is to consider an arbitrary smooth function of the scalar torsion to construct the action
\begin{align}
S_{f(T)} = \int f(T)e\, d^4x \, .\label{f(T)}
\end{align}
This theory is called ``$f(T)$ gravity" and it has numerous and interesting applications, for example in cosmology (See \cite{Cai:2015emx} for a comprehensive review of those models). One important feature of this theory is that meanwhile TEGR is an equivalent theory to GR, $f(T)$ does not produce the same field equations as $f(R)$ gravity (due to the relationship \eqref{RTB} ) and therefore one needs to consider a generalisation of \eqref{f(T)} from $f(T)\rightarrow f(T,B)$ to find the teleparallel equivalent to $f(R)$ gravity as discussed in  \cite{Bahamonde:2015zma}.
Starting from the action \eqref{f(T)}, the field equations read
\begin{align}
  4e\Big[f_{TT}(\partial_{\mu}T)\Big]S_{\nu}{}^{\mu\lambda}+
  4e^{a}_{\nu}\partial_{\mu}(e S_{a}{}^{\mu\lambda})f_{T}\nonumber\\
  -4ef_{T}T^{\sigma}{}_{\mu \nu}S_{\sigma}{}^{\lambda\mu}-
  ef\delta_{\nu}^{\lambda} = 16\pi e \Theta_{\nu}^{\lambda} \,,
  \label{fTT}
\end{align}
where the energy-momentum tensor is defined as follows
\begin{align}
  \Theta^{\lambda}_{a}=\frac{1}{e} \frac{\delta (eL_m)}{\delta e^{a}_{\lambda}} \,.
\end{align}
With this considerations in mind, let us start our discussion on black holes in $f(T)$ gravity.

\section{Black hole in $f(T)$ gravity\label{secFR}}
The  metric for a spherically symmetric black hole with mass $M$ in $f(T)$ gravity is given by~\cite{Paliathanasis:2014iva}
\begin{multline}\label{1}
ds^{2}= -A\,dt^2+\frac{dr^{2}}{c_{3}^{2}A}+r^2(d\theta^2+\sin^2\theta d\phi^2),
\end{multline}
where
\begin{eqnarray}
A&\equiv& \frac{2c_{1}r^{2}}{3c_{3}}-\frac{2c_{5}}{c_{3}r}=\frac{2Xr^{2}}{3}-\frac{2C_{5}}{r},\\
\text{where,~~~~~}c_{5}&\equiv& c_{1}c_{4}-c_{2}c_{3},\\
\text{and,~~~~~}X&\equiv&\frac{c_{1}}{c_{3}};~~~~~C_{5}\equiv\frac{c_{5}}{c_{3}}.
\end{eqnarray}
Here, all $c_{1},c_{2},c_{3},c_{4}$ and $c_{5}$ are constants. The horizon is given by
\begin{equation}\label{3a}
r_h=\Big(\frac{3c_{5}}{c_{1}}\Big)^{1/3}=\Big(\frac{3C_{5}}{X}\Big)^{1/3},
\end{equation}
where we have introduced the new constants $C_{5}=c_{5}/c_{3}$ and $X=c_{1}/c_{3}$ which will turn very useful in the study of the dynamical system. To ensure that $r_h>0$, $C_{5}$ and $X$ must have the same sign: $C_{5}/X>0$. Since $A$ must be positive at spatial infinity, we must have $X>0$ resulting in $C_{5}>0$.
Upon performing the coordinate transformation
\begin{equation}\label{2m}
    t = c_{3}t',
\end{equation}
we bring the metric~\eqref{1} to the following form where $\al (r)=c_{3}^2A(r)$
\begin{equation}\label{3m}
ds^{2}= -\al (r) dt'^2+\frac{dr^{2}}{\al (r)}+r^2(d\theta^2+\sin^2\theta d\phi^2).
\end{equation}
This is precisely the general form of metric used in Ref.~\cite{Ahmed:2015tyi}  {where accretions of samples of $f(R)$ black holes were investigated, among which we find the solution
\begin{equation}\label{mfr}
\al(r)\equiv 1-\frac{2M}{r}+\beta r-\frac{\Lambda r^{2}}{3}.
\end{equation}
This will serve in Sec.~\ref{ftr} as a tool for comparing accretion onto the $f(T)$ black hole~\eqref{1} with that onto the $f(R)$ black hole~\eqref{mfr}.}

Metric~\eqref{1} being equivalent to~\eqref{3m}, all general equations expressed in terms of $\al$, which were derived in Ref.~\cite{Ahmed:2015tyi}, are thus applicable to our present investigation upon replacing $\al$ by $c_{3}^2A$. However, because of their importance, we will outline their derivations below.

\section{General equations for spherical accretion\label{secGE}}
Let $n$ be the baryon number density in the fluid rest frame and $u^{\mu}=dx^{\mu}/d\tau$ be the four velocity of the fluid where $\tau$ is the proper time. We define the particle flux or current density by $J^{\mu}=n u^{\mu}$ where $n$ is the particle density. From the  particle conservation law, we have that the divergence of current density is conserved, i.e.
\begin{equation}\label{6a}
\nabla_{\mu}J^{\mu}=\nabla_{\mu}(n u^{\mu})=0,
\end{equation}
where $\nabla_{\mu}$ is the covariant derivative. On the other hand, energy-momentum tensor is explicitly given by
\begin{eqnarray}\label{7}
\Theta^{\mu \nu}&=&(\epsilon+p)u^{\mu}u^{\nu}+p g^{\mu \nu},
\end{eqnarray}
where $\epsilon$ denotes the energy density and $p$ is the pressure. We assume that the fluid is radially flowing in the equatorial plane $(\theta=\pi/2)$, therefore $u^{\theta}=0$ and $u^{\phi}=0$. For the sake of simplicity,  we set $u^r=u$. Using the normalization condition $u^{\mu}u_{\mu}=-1$ and~\eqref{1}, we obtain,
\begin{equation}\label{9}
u_{t}= -\frac{\sqrt{c_{3}^{2}A+u^2}}{c_{3}}.
\end{equation}
On the equatorial plane $(\theta=\pi/2)$, the continuity equation~(\ref{6a}) yields
\begin{eqnarray}\label{10}
\nabla_{\mu}(n u^{\mu})&=& \frac{1}{\sqrt{-g}}\partial_{\mu}(\sqrt{-g}n u^{\mu})\nonumber \\&=&
\frac{1}{r^{2}}\partial_{r}(r^{2}n u)=0.
\end{eqnarray}
or, upon integrating,
\begin{equation}\label{17}
r^{2}n u=C_{1},
\end{equation}
where $C_{1}$ is a constant of integration.
The thermodynamics of simple fluids is described by ~\cite{Rezzolla}
\begin{equation}\label{t1}
dp=n(dh-Tds),\quad d\epsilon=hdn+nTds,
\end{equation}
where $T$ is the temperature, $s$ is the specific entropy and
\begin{equation}\label{ent}
h=\frac{\epsilon+p}{n},
\end{equation}
is the specific enthalpy.
On the other hand, a  theorem of relativistic hydrodynamics~\cite{Rezzolla} states that the scalar $hu_{\mu}\xi^{\mu}$ is conserved along the trajectories of the fluid,
\begin{equation}\label{t3}
u^{\nu}\nabla_{\nu}(hu_{\mu}\xi^{\mu})=0,
\end{equation}
where $\xi^{\mu}$ is a Killing vector of spacetime. Consider the timelike Killing vector $\xi^{\mu}=(1,0,0,0)$ of the metric~\eqref{1}, we obtain
\begin{equation}\label{t4}
\partial_{r}(hu_{t})=0\quad\text{or}\quad h\sqrt{c_{3}^{2}A+u^2}=C_2,
\end{equation}
where $C_2$ is a constant of integration. It is easy to show that the specific entropy is conserved along the fluidlines: $u^{\mu}\nabla_{\mu}s=0$. In fact, if we rewrite energy-momentum tensor $\Theta^{\mu \nu}$~\eqref{7} as $nhu^{\mu}u^{\nu}+(nh-e)g^{\mu\nu}$~\cite{Ahmed:2015tyi}, then project the conservation formula of $\Theta^{\mu \nu}$ onto $u^{\mu}$, we obtain:
\begin{align}\label{t2}
u_{\nu}\nabla_{\mu}\Theta^{\mu\nu}=&\ u_{\nu}\nabla_{\mu}[nhu^{\mu}u^{\nu}+(nh-e)g^{\mu\nu}]\nn\\
\quad =\ &u^{\mu}(h\nabla_{\mu}n-\nabla_{\mu}e)=-nTu^{\mu}\nabla_{\mu}s=0.
\end{align}
In the special case we are considering in this work where the fluid motion is radial, stationary (no dependence on time), and it conserves the spherical symmetry of the black hole, the latter equation reduces to $\partial_rs=0$ everywhere, that is, $s\equiv \text{const.}$. Thus, the motion of the fluid is isentropic and equations~\eqref{t1} reduce to
\begin{equation}\label{t1b}
dp=ndh,\quad d\epsilon =hdn.
\end{equation}
Equations~\eqref{17}, \eqref{t4}, and~\eqref{t1b} are the main equations that we will use  to analyze the flow.
Since $s$ is constant, this reduces the canonical form of the equation of state (EOS) of a simple fluid $e=e(n,s)$ to the barotropic form
\begin{equation}\label{b1}
    \epsilon=F(n).
\end{equation}
From the second equation~\eqref{t1b},  we have $h=d\epsilon/dn$ which yields
\begin{equation}\label{b2}
    h=F'(n),
\end{equation}
where the prime denotes differentiation with respect to $n$. Now, the first equation~\eqref{t1b} yields $p'=nh'$, with $h=F'$, we obtain
\begin{equation}\label{b3}
p'=nF'',
\end{equation}
which can be integrated by parts to derive
\begin{equation}\label{b4}
p=nF'-F.
\end{equation}
We have that an EOS of the form $p=G(n)$ is not independent of an EOS of the form $\epsilon=F(n)$. The relation between $F$ and $G$ can be derived upon integrating the differential equation
\begin{equation}\label{b5}
nF'(n)-F(n)=G(n).
\end{equation}
The local three-dimensional speed of sound $a$ is defined by $a^2=(\partial p/\partial \epsilon)_s$. Since the entropy $s$ is constant, this reduces to $a^2=dp/d\epsilon$. Using~\eqref{t1b}, we derive
\begin{equation}\label{19}
a^2=\frac{dp}{d\epsilon}=\frac{ndh}{hdn}\Rightarrow \frac{dh}{h}=a^{2}\frac{dn}{n}.
\end{equation}
Using~\eqref{b2}, this reduces to
\begin{equation}\label{19b}
a^2=\frac{ndh}{hdn}=\frac{n}{F'}F''=n(\ln F')'.
\end{equation}
Since the motion is radial in the plane $\ta=\pi/2$, we have $d\ta=d\phi=0$ and the metric~\eqref{1} implies the decomposition
\begin{equation*}
    ds^2=-(\sqrt{A}dt)^2+(dr/c_{3}\sqrt{A})^2
\end{equation*}
The ordinary three-dimensional speed $v$ is defined by $v\equiv \frac{dr/\sqrt{A}}{c_{3}\sqrt{A}dt}$ and yields
\begin{equation}\label{v}
v^2=\Big(\frac{u}{c_{3}Au^t}\Big)^2=\frac{u^2}{c_{3}^{2}A+u^2},
\end{equation}
where we have used $u^r=u=dr/d\tau$, $u^t=dt/d\tau$, $u_t=-Au^t$, and~\eqref{9}. This implies
\begin{equation}\label{v2}
u^2=\frac{c_{3}^{2}Av^2}{1-v^2}\quad \text{ and }\quad (u_t)^2=\frac{A^2}{1-v^2},
\end{equation}
and~\eqref{17} becomes
\begin{equation}\label{v3}
\frac{r^4n^2c_{3}^2Av^2}{1-v^2}=C_1^2.
\end{equation}
These results will be used in the following Hamiltonian analysis.
\section{Hamiltonian systems\label{secHS}}
We have derived two integrals of motion ($C_1,C_2$) given in~\eqref{17} and~\eqref{t4}. Let $\mathcal{H}$ be the square of the lhs of~\eqref{t4}:
\begin{equation}\label{h1}
\mathcal{H}=h^2(c_{3}^2A+u^2).
\end{equation}
Using~\eqref{v2} the Hamiltonian~\eqref{h1} of the dynamical system reads
\begin{equation}\label{h3}
\mathcal{H}(r,v)=\frac{h(r,v)^2c_{3}^2A}{1-v^2},
\end{equation}
as derived in Ref.~\cite{Ahmed:2015tyi} where $f$ has been replaced by $c_{3}^2A$. We can absorb the constant $c_{3}^2$ into a redefiniton of the Hamiltonian, however, we will do that in a further step of derivation.
\subsection{Sonic points}
With $\mathcal{H}$ given by~\eqref{h3}, the dynamical system reads
\begin{equation}\label{h4}
\dot{r}=\mathcal{H}_{,v}\,,  \quad\quad \dot{v}=-\mathcal{H}_{,r}.
\end{equation}
(here the dot denotes the $\bar{t}$ derivative). Evaluating the rhs's we find
\begin{align}
\label{h5a}&\mathcal{H}_{,v}=\frac{2c_{3}^2h^2Av}{(1-v^2)^2}\Big[1+\frac{1-v^2}{v}~(\ln h)_{,v}\Big],\\
\label{h5b}&\mathcal{H}_{,r}=\frac{c_{3}^2h^2}{1-v^2}\big[A_{,r}+2A~(\ln h)_{,r}\big].
\end{align}
Following the same approach as in Ref.~\cite{Ahmed:2015tyi},  we arrive at
\begin{align}
\label{h9a}&\dot{r}=\frac{2c_{3}^2h^2A}{v(1-v^2)^2}~(v^2-a^2),\\
\label{h9b}&\dot{v}=-\frac{c_{3}^2h^2}{r(1-v^2)}[rA_{,r}(1-a^2)-4Aa^2].
\end{align}
For the CP, the rhs's vanish if the conditions
\begin{equation}\label{cp1}
v_c^2=a_c^2\quad\text{ and }\quad r_c(1-a_c^2)A_{c,r_c}=4A_ca_c^2,
\end{equation}
hold.  {Here $A_c\equiv A(r_c)$ and $A_{c,r_c}\equiv A_{,r}|_{r=r_c}$}. They  lead to
\begin{equation}\label{cp2}
    a_c^2=\frac{r_cA_{c,r_c}}{r_cA_{c,r_c}+4A_c}.
\end{equation}
 {If solutions to the system of equations~\eqref{cp1} exist, we rewrite the constant $C_1^2$ in~\eqref{v3} as
\begin{equation}\label{b9}
    C_1^2=r_c^4n_c^2c_{3}^2v_c^2~\frac{A_c}{1-v_c^2}=\frac{r_c^5n_c^2c_{3}^2A_{c,r_c}}{4},
\end{equation}
where we have used the second equation in~\eqref{cp1}. Using this in~\eqref{v3} we obtain the result
\begin{equation}\label{b10}
\Big(\frac{n}{n_c}\Big)^{2}=\frac{r_c^5A_{c,r_c}}{4}~\frac{1-v^2}{r^4Av^2}.
\end{equation}
If no solution to~\eqref{cp1}, we can keep~\eqref{v3} as it is or introduce any point ($r_0,v_0$) from the phase portrait to obtain
\begin{equation}\label{b10b}
\hspace{-2mm}n^2=\Big(\frac{C_1}{c_3}\Big)^2~\frac{1-v^2}{r^4Av^2}\quad\text{or}\quad \Big(\frac{n}{n_0}\Big)^{2}=\frac{r_0^4A_0v_0^2}{1-v_0^2}~\frac{1-v^2}{r^4Av^2}.
\end{equation}
}
The above dynamical system allows to perform the analysis of the fluids that we are considering.
\section{Isothermal test fluids\label{secITS}}
Isothermal flow is often referred to the fluid flowing at a constant temperature. In this section we find the general solution of  the isothermal EOS of the form $p=k\epsilon$, that is of the form $p=kF(n)$~\eqref{b1} with $G(n)=kF(n)$~\eqref{b5}. Here $k$ is the state parameter such that $(0<k\leq1)$~\cite{t1}. The differential equation~\eqref{b5} reads
\begin{equation}\label{b6}
nF'(n)-F(n)=kF(n),
\end{equation}
 {yielding
\begin{equation}\label{b7}
    \epsilon=F=\frac{\epsilon_c}{n_c^{k+1}}\,n^{k+1}=\frac{\epsilon_0}{n_0^{k+1}}\,n^{k+1},
\end{equation}}%
where we have chosen the constant of integration\footnote{This constant, $\epsilon_c/n_c^{k+1}$, in~\eqref{b7} can be  chosen as $\epsilon_{\infty}/n_{\infty}^{k+1}$ or $\epsilon_0/n_0^{k+1}$ where ($\epsilon_0,n_0$) are energy density and  number density.} so that~\eqref{ent} and~\eqref{b2} lead to the same expression for $h$
\begin{equation}\label{b8}
    h=\frac{(k+1)\epsilon_c}{n_c^{k+1}}\,n^{k}=\frac{(k+1)\epsilon_c}{n_c}\Big(\frac{n}{n_c}\Big)^{k}.
\end{equation}
 {Using~\eqref{b10} or~\eqref{b10b},  we obtain}
\begin{equation}\label{b11}
    h^2\propto \Big(\frac{1-v^2}{v^2r^4A}\Big)^{k}
\end{equation}
and
\begin{align}\label{nds}
&\mathcal{H}(r,v)=\frac{A^{1-k}}{(1-v^2)^{1-k} v^{2 k} r^{4 k}},
\end{align}
where all the constant factors have been absorbed into the  redefinition of the time $\bar{t}$ and the Hamiltonian $\mathcal{H}$.
Now we will analyze the behavior of the fluid by taking different cases for the state parameter $k$. For instance, we have $k=1$ (ultra-stiff fluid), $k=1/2$ (ultra-relativistic fluid), $k=1/3$ (radiation fluid) and $k=1/4$ (sub-relativistic fluid). In  the case of the metric~\eqref{1}, Eq. (\ref{cp2}) reduces to
\begin{equation}\label{cp3a}
    k=\frac{2c_{1}r_c^3 + 3c_{5}}{6c_{1}r_c^3 - 9c_{5}}=\frac{2Xr_c^3 + 3C_{5}}{6Xr_c^3 - 9C_{5}}
\end{equation}
and yields
\begin{equation}\label{cprh}
r_c = \Big(\frac{3k+1}{2(3k-1)}\Big)^{1/3}r_h ,
\end{equation}
where $r_h$ is given by~\eqref{3a}. It is easy to see that, in order to have $r_c>r_h>0$, we must have $C_{5}/X>0$ and $1/3<k<1$. This fixes the values of $k$ that yield a critical flow with the presence of a CP given by~\eqref{cprh} and $v_c^2=k$.
In Ref.~\cite{Ahmed:2015tyi} we have shown that if the flow approaches the horizon with a vanishing three-dimensional speed, the pressure must diverge as
\begin{equation}\label{pr2}
    p\sim (r-r_h)^{-\frac{k+1}{2k}},
\end{equation}
if $A(r)=0$ has a single root.
\subsection{Solution for ultra-stiff fluid (${\pmb k=1}$)}
The equation of state for the ultra-stiff fluids is $p = k\epsilon$ i.e. the value of state parameter is defined as $k=1$. The Hamiltonian~\eqref{nds} reduces to
\begin{equation}\label{k11}
\mathcal{H}=\frac{1}{v^2r^4}.
\end{equation}
\begin{figure}[!htb]
\centering
\includegraphics[width=0.45\textwidth]{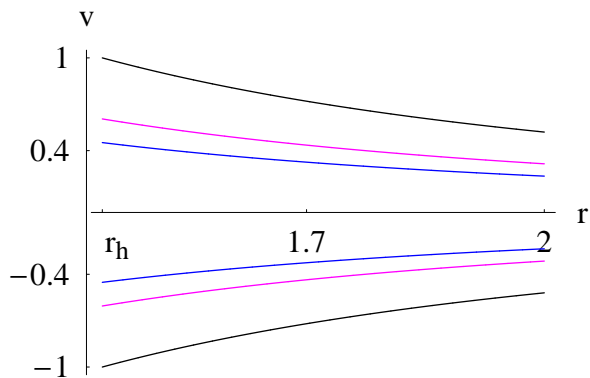}\\
\caption{Case $k=1$. Plot of $\mathcal{H}$~\eqref{k11} for $C_{5}=X=1$. The event horizon~\eqref{3a} is at $r_h =3^{1/3}$. Black plot: the solution curve corresponding to $\mathcal{H}=\mathcal{H}_{\text{min}}=r_h^{-4}$.  The magenta and blue plots correspond to $\mathcal{H}>\mathcal{H}_{\text{min}}$.}\label{Fig1}
\end{figure}
From~\eqref{k11} we see that, for physical flows ($|v|<1$), the lower value of $\mathcal{H}$ is $\mathcal{H}_{\text{min}}=1/r_h^4$: $\mathcal{H}>\mathcal{H}_{\text{min}}$. As shown in Fig.~\ref{Fig1}, physical flows are represented by the curves sandwiched by the two black curves, which are contour plots of $\mathcal{H}(r,v)=\mathcal{H}_{\text{min}}$. The upper curves where $v>0$ correspond to fluid outflow or particle emission and the lower curves where $v<0$ correspond to fluid accretion.
From~\eqref{k11} we see that for the global solutions, shown in Fig.~\ref{Fig1}, which are the only existing solution for $k=1$, the speed $v$ behaves asymptotically as $v\sim 1/r^2$. Using this and the fact that $A\sim r^2$ in~\eqref{v3}, we obtain $n\sim 1/r$.
\subsection{Solution for ultra-relativistic fluid (${\pmb k=1/2}$)}
\begin{figure}[!htb]
\centering
\includegraphics[width=0.45\textwidth]{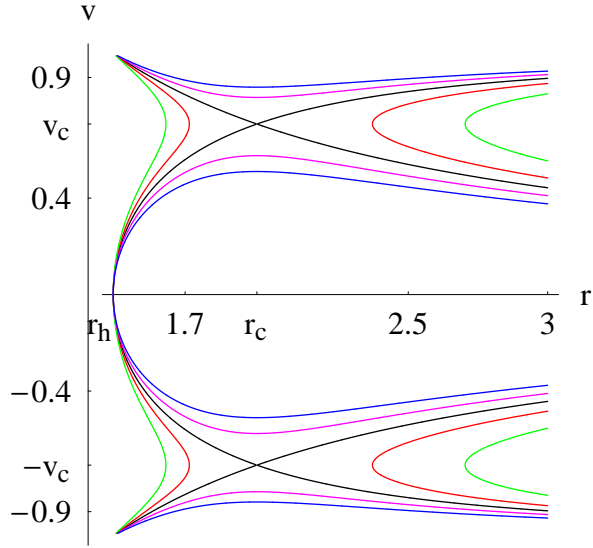}\\
\caption{Case $k=1/2$. Plot of $\mathcal{H}$~\eqref{k121} for $C_{5}=X=1$. The event horizon~\eqref{3a} is at $r_h =3^{1/3}$ and $r_c=5r_h/2$. Black plot: the solution curve through the saddle CPs $(r_{c},v_c)$ and $(r_{c},-v_c)$ corresponding to $\mathcal{H}=\mathcal{H}_c\simeq 0.646209$. The magenta and blue plots correspond to $\mathcal{H}>\mathcal{H}_c$ and the red and green plots to $\mathcal{H}<\mathcal{H}_c$.}\label{Fig2}
\end{figure}

Ultra-relativistic fluids are those fluids whose isotropic pressure is less than the energy density.
In this case, the equation of state is defined as $p=\frac{\epsilon}{2}$ yielding $k=1/2$.
Using this expression in~\eqref{cprh} reduces to $r_c=5r_h/2$. Thus, we have two CPs given by
\begin{align}\label{k12cp}
&r_c=5r_h/2,\quad v_c=\sqrt{1/2},\nn\\
&r_c=5r_h/2,\quad v_c=-\sqrt{1/2}.
\end{align}
The Hamiltonian~\eqref{nds} takes the simple form
\begin{equation}\label{k121}
\mathcal{H}=\frac{\sqrt{A}}{r^2|v|\sqrt{1-v^2}}.
\end{equation}
For some given value of $\mathcal{H}=\mathcal{H}_0$, Eq.~\eqref{k121} can be solved for $v^2$. Another way to represent the flow is to use contour plots as shown in Fig.~\ref{Fig2}. For the global solutions depicted in the figure, the speed $v$ has two different asymptotic behaviors. Since $\mathcal{H}$ retains the same constant value and $A\sim r^2$, we have either (a) $v\to 0$ as $v\sim cst/r$ or (b) $v\to 1$ such that $r^2(1-v^2)\sim cst$ yielding $v\sim 1-cst/(2r^2)$. Using these in~\eqref{v3}, we obtain (a) $n\sim 1/r^2$ and (b) $n\sim 1/r^4$.
The plot shows three main types of fluid motion:
\begin{enumerate}
  \item Purely supersonic accretion ($v<-v_c$), which ends inside the horizon, or purely supersonic outflow ($v>v_c$);
  \item Purely subsonic accretion followed by subsonic flowout, this is the case of the branches of the blue and magenta solution curves corresponding to $-v_c<v<v_c$. Notice that for this motion the fluid reaches the horizon, $A(r_h)=0$, with a vanishing speed ensuring that the Hamiltonian~\eqref{k121} remains constant.
      The critical black solution curve reveals two types of motions: if we assume that $dv/dr$ is continuous at the CPs;
  \item \begin{description}
          \item [a.] Supersonic accretion until ($r_c,-v_c$), followed by a subsonic accretion until ($r_h,0$), where the speed vanishes, then a subsonic flowout until ($r_c,v_c$), followed by a supersonic flowout;
          \item [b.] Subsonic accretion followed by a supersonic accretion which ends inside the horizon. In the upper plot, we have a supersonic outflow followed by a subsonic motion.
        \end{description}
\end{enumerate}
The fluid flow in Type (3) from ($r_c,-v_c$) to ($r_c,v_c$) describes a heteroclinic orbit that passes through two different saddle CPs: ($r_c,-v_c$) and ($r_c,v_c$). It is easy to show that the solution curve from ($r_c,-v_c$) to ($r_c,v_c$) reaches ($r_c,v_c$) as $\bar{t}\to -\infty$, and the curve from ($r_c,v_c$) to ($r_c,-v_c$) reaches ($r_c,-v_c$) as $\bar{t}\to +\infty$; we can change the signs of these two limits upon performing the transformation $\bar{t}\to -\bar{t}$ and $\mathcal{H}\to -\mathcal{H}$.
The flowout of the fluid, which starts at the horizon, is caused by the high pressure of the fluid, which diverges there~\eqref{pr2}: The fluid under effects of its own pressure flows back to spatial infinity.
Not all solution curves shown in Fig.~\ref{Fig2} are physical. Recall that the analysis made in this paper considers the fluid elements as test particles not modifying the geometry of the $f(T)$ black hole. It is thus assumed that the accretion does not modify the mass of the black hole nor its other intrinsic properties. The flow being non-geodesic, however still obeys the simple rule that if $r$ increases, $v$ must be positive, and if $r$ decreases, $v$ must be negative. For instance, for $v>0$, we see from Fig.~\ref{Fig2} that the red plot has two branches. Consider the branch on the right of the vertical line $r=r_c$. The flow along the segment of that branch along which $v$ increases and $r$ decreases is unphysical, for this is neither an accretion nor a flowout.
\subsection{Solutions for radiation fluid (${\pmb k=1/3}$) and sub-relativistic fluid (${\pmb k=1/4}$)}
\begin{figure}[!htb]
\centering
\includegraphics[width=0.45\textwidth]{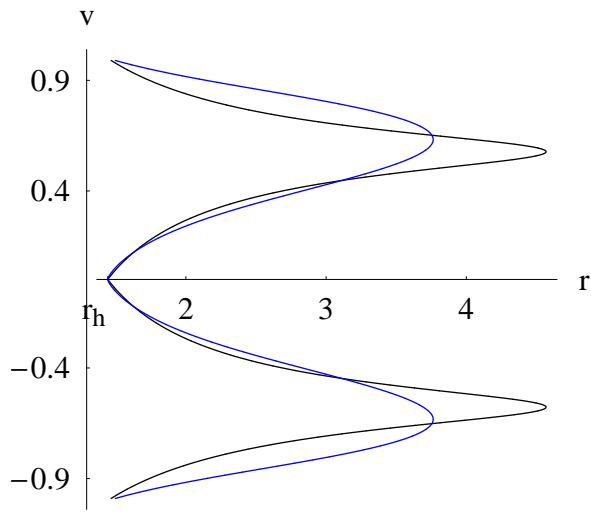}\\
\caption{Case $k\leq 1/3$. Black plot: contour plots of $\mathcal{H}$~\eqref{k13p} for $C_{5}=X=1$ (k=1/3). Blue plot: contour plots of $\mathcal{H}$~\eqref{k14p} for $C_{5}=X=1$ (k=1/4). The event horizon~\eqref{3a} is at $r_h =3^{1/3}$.}\label{Fig3}
\end{figure}

Radiation fluids ($k=1/3$) are the fluid which absorbs the radiation emitted by the black hole. It is the most interesting case in astrophysics and sub-relativistic fluids ($k=1/4$) are those fluids whose energy density exceeds their isotropic pressure. The Hamiltonian~\eqref{nds} for these fluids takes the following expressions, respectively
\begin{align}
\label{k13p}&\mathcal{H}=\frac{A^{2/3}}{r^{4/3}|v|^{2/3}(1-v^2)^{2/3}}& & (k=1/3),\\
\label{k14p}&\mathcal{H}=\frac{A^{3/4}}{r\sqrt{|v|}(1-v^2)^{3/4}}& & (k=1/4).
\end{align}
As we concluded earlier in this section, there is no critical flow for these fluids and for all fluid cases where $k\leq 1/3$; rather, simple fluid flow characterizes this class of fluids. Moreover, the fluid flow for this class of fluids is not global, in that, it does not extend to spatial infinity except in the case $k=1/3$ where the flow can be global and non-global. This conclusion can be derived from~\eqref{nds} as follows. If the flow is global, $v$ behaves asymptotically as
\begin{equation}
v\simeq v_0r^{-\al}+v_{\infty},
\end{equation}
where $\al >0$, $v_0$, and $|v_{\infty}|\leq 1$ are constants. If we assume that the flow is global, that is, $r$ may go to infinity, the Hamiltonian~\eqref{nds} behaves in the limit $r\to\infty$ as
\begin{align}
&\mathcal{H}\propto r^{2(1-3k+k\al)}& & (v_{\infty}=0),\nn\\
\label{k13p2}&\mathcal{H}\propto r^{2(1-3k)}& & (0<|v_{\infty}|<1),\\
&\mathcal{H}\propto r^{2(1-3k)+(1-k)\al}& & (|v_{\infty}|=1).\nn
\end{align}
Thus, in the case $k< 1/3$, the Hamiltonian diverges at spatial infinity. Since the Hamiltonian is a constant of motion, the assumption that $r$ goes to infinity is not valid. For $k=1/3$ global flow is possible, as we shall justify below, however non-global flow is also realizable. Fig.~\ref{Fig3} depicts typical non-global fluid flows for this class of fluids where $k\leq 1/3$. Let $r_{\text{rm}}$ be the $r$ coordinate of the rightmost point on the solution curve. We observe:
\begin{enumerate}
  \item (Generally supersonic) accretion from $r_{\text{rm}}$ that crosses the horizon with the speed of light. Such flow is possible if a fluid source is available at $r_{\text{rm}}$ that injects fluid particles with a nonvanishing speed;
  \item (Almost subsonic) accretion from $r_{\text{rm}}$ that reaches the horizon with a vanishing speed, followed by a (almost subsonic) flowout back to $r_{\text{rm}}$. Such flow could be made possible if a source-sink system is available at $r_{\text{rm}}$;
  \item (Generally supersonic) flowout that emanates from the horizon with the speed of light and reaches $r_{\text{rm}}$ with a nonvanishing speed. Such flow is possible if a sink is available at $r_{\text{rm}}$.
\end{enumerate}
\begin{figure}[!htb]
\centering
\includegraphics[width=0.45\textwidth]{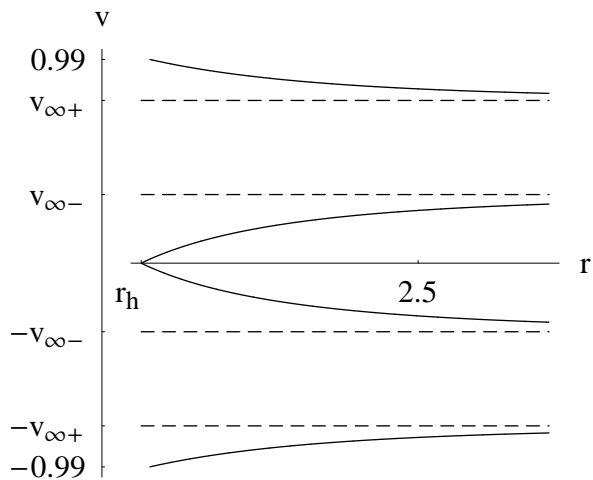}\\
\caption{Case $k= 1/3$. Contour plots of the Hamiltonian~\eqref{k13p} with value~\eqref{k13p3} $\mathcal{H}=2.25 (2X/3)^{2/3}>(3X^2)^{1/3}$ for $C_{5}=X=1$ showing a global solution. The event horizon~\eqref{3a} is at $r_h =3^{1/3}$, $v_{\infty-}=1/3$, and $v_{\infty+}=\sqrt{(17-\sqrt{33})/18}\simeq 0.79076$.}\label{Fig4}
\end{figure}
Now, if $k=1/3$ and $0<|v_{\infty}|<1$, the Hamiltonian~\eqref{k13p2} has a finite limit as $r\to\infty$, so global flow is possible. To achieve it, that is, to determine such global flow solutions, notice that the value of the Hamiltonian~\eqref{k13p} in this case is
\begin{equation}\label{k13p3}
\mathcal{H}=\frac{1}{|v_{\infty}|^{2/3}(1-v_{\infty}^2)^{2/3}}\Big(\frac{2X}{3}\Big)^{2/3}\qquad (0<|v_{\infty}|<1).
\end{equation}
Since $0<|v_{\infty}|<1$, we have $0<|v_{\infty}|^{2/3}(1-v_{\infty}^2)^{2/3}\leq 4^{1/3}/3$. Hence, to have such global flow solutions, we must restrict the value of the Hamiltonian by
\begin{equation}
\mathcal{H}\geq \frac{3}{4^{1/3}}\Big(\frac{2X}{3}\Big)^{2/3}=(3X^2)^{1/3}.
\end{equation}
Non-global solutions correspond to $0<\mathcal{H}<(3X^2)^{1/3}$. Notice also that for a given value of $\mathcal{H}>(3X^2)^{1/3}$, there are two possible values of $|v_{\infty}|$, denoted by ($v_{\infty-},v_{\infty+}$), such that $v^2_{\infty-}<1/3$ and $v^2_{\infty-}>1/3$; for $\mathcal{H}=(3X^2)^{1/3}$ we have $v^2_{\infty-}=v^2_{\infty+}=1/3$. It is easy to show that for $v_{\infty}=v_{\infty-}$, $v_0<0$ and that for $v_{\infty}=v_{\infty+}$, $v_0>0$, as shown in Fig.~\ref{Fig4}. Figure~\ref{Fig4} depicts a typical global fluid flow for $k= 1/3$. We observe three types of flow:
\begin{enumerate}
  \item (Supersonic) accretion with an initial velocity $-v_{\infty+}$ that crosses the horizon with the speed of light;
  \item (Subsonic) accretion with an initial velocity $-v_{\infty-}$ that reaches the horizon with a vanishing speed, followed by a (subsonic) flowout that reaches spatial infinity with the same speed $v_{\infty-}$;
  \item (Supersonic) flowout that emanates from the horizon with the speed of light and reaches spatial infinity with a speed $v_{\infty+}$.
\end{enumerate}
For the global flow, we determine the particle density $n$ as follows. Equation~\eqref{k13p3} with $\mathcal{H}$ given by the rhs of~\eqref{k13p} yields
\begin{equation}\label{n1}
A = \frac{2X}{3}~\frac{|v|(1-v^2)}{|v_{\infty}|(1-v_{\infty}^2)}~r^2.
\end{equation}
Substituting this in~\eqref{v3} we obtain
\begin{equation}\label{n2}
n^2=\frac{N^2}{|v|^3r^6},
\end{equation}
where all constants ($X,c_{3},v_{\infty}$) have been grouped or absorbed into the new constant $N^2$. Since asymptotically $|v|\to v_{\infty\pm}$, which is a nonzero constant, $n\sim r^{-3}$.

\subsection{Accretions in  $f(T)$ and $f(R)$ gravities\label{ftr}}
We draw a comparison between accretions in $f(T)$ and $f(R)$ gravities. For that end we select from $f(R)$ gravity black holes a similar solution~\eqref{mfr} to the one considered here~\eqref{1}, that is, an anti-de Sitter-like $f(R)$ black hole~\cite{Ahmed:2015tyi}. The following enumeration shows similarities and differences.
\begin{enumerate}
  \item The accretion of an isothermal perfect fluid with $k=1$ is characterized by the presence of global solutions, which are the only existing solutions with no CPs. The speed $v$ and the particle density $n$ behave asymptotically as $v\sim 0$ and $n\sim 1/r$ for both gravities;
  \item If the isothermal perfect fluid has $k=1/2$, the accretion is characterized by the presence of two CPs and critical flow for both gravities. For the global solutions we have either $v\sim 0$ and $n\sim 1/r^2$ or $v\sim 1$ and $n\sim 1/r^4$;
  \item \begin{description}
          \item [a.] For $f(T)$ gravity the accretion of an isothermal perfect fluid with $k=1/3$ has no CP nor critical flow while for $f(R)$ gravity the fluid flow has two CPs. For the global solutions of both gravities $v\sim cst$, where $cst$ may assume any value between 0 and 1, and $n\sim 1/r^3$;
          \item [b.] For $k<1/3$, the accretion onto an $f(T)$ gravity black hole is again noncritical, with no CP, while that onto an $f(R)$ gravity black hole may have four CPs, as was shown in Ref.~\cite{Ahmed:2015tyi} for the isothermal perfect fluid with $k=1/4$. For both gravities there are no global solutions.
        \end{description}
\end{enumerate}

This, however, is just a qualitative comparison. First of all notice that the black hole~\eqref{mfr} of the $f(R)$ gravity reduces to that of GR and the theory itself reduces to GR, $f(R)=R+\La$, if the $f(R)$-parameter $\bt=0$. This is not the case with the black hole~\eqref{1} of the $f(T)$ gravity which does not reduce to any of the known GR black holes no matter how the $f(T)$-parameters, ($X,C_5$), are chosen.

A deeper investigation should focus on the evaluation of the rates of accretion and efficiencies of the outgoing spectra for different black holes and different gravity theories.

The efficiency of conversion of gravitational (potential) energy into radiation is one of the open problems of radial accretion onto a black hole, this is if one assumes, as most workers concluded, that the infall velocity scales almost as the free fall velocity (the case of Fig.~\ref{Fig1} or the case of the critical subsonic accretion followed by a supersonic accretion of Fig.~\ref{Fig2}). This efficiency problem becomes more involved if we consider the critical accretion of Fig.~\ref{Fig2} along the branch where $v$ vanishes as $r\to r_h$ or accretions along the blue and magenta branches of the same figure. Here the three velocity has a deceleration phase from $r_c$ to $r_h$ and it does not scale as a free fall velocity. This is our main discovery in this work and in~\cite{Ahmed:2015tyi}. The deceleration of the fluid increases by far the conversion efficiency; moreover, the efficiency is roughly proportional to $n^2$~\cite{basic}, which diverges by~\eqref{b10b} as $r\to r_h$.

All that is out of the scope of this work and could be the aim and task of subsequent works. In a first step one may consider the simplest cases of the $f(T)=T$ [$f(R)=R$ or GR] gravity theory. We believe that, when all these tasks are performed (most likely numerically), the result that will be at hand will confirm the equivalency of these gravity theories.

\section{Polytropic test fluids\label{secptf}}
The polytropic equation of state is
\begin{equation}\label{p1}
p=G(n)=Kn^{\gamma},
\end{equation}
where $K$ and $\gamma$ are constants. For ordinary matter, one generally works with the constraint $\ga>1$. Inserting~\eqref{p1} into the differential equation~\eqref{b5}, it is easy to establish~\cite{Ahmed:2015tyi} the following expressions of the specific enthalpy
\begin{equation}\label{pl2}
  h=m+\frac{K\ga n^{\ga-1}}{\ga-1},
\end{equation}
by integration, and the three-dimensional speed of sound from~\eqref{19b}
\begin{equation}\label{pl3}
   a^2=\frac{(\ga -1)Y}{m(\ga -1)+Y}\qquad (Y\equiv K\ga n^{\ga -1}),
\end{equation}
where we have introduced the baryonic mass $m$. Since $\ga >1$, this implies $a^2<\ga -1$ and, particularly, $v_c^2<\ga -1$.
\begin{figure*}[!htb]
\centering
\includegraphics[width=0.32\textwidth]{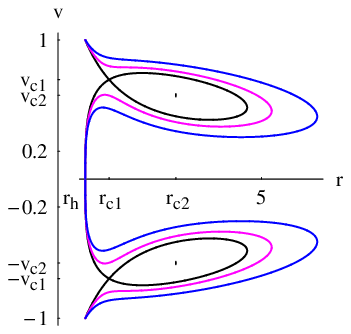} \includegraphics[width=0.32\textwidth]{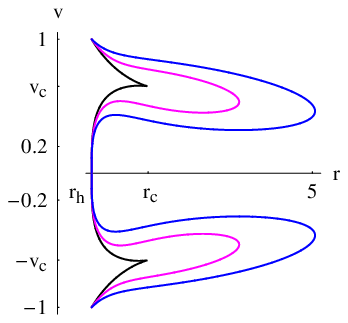} \includegraphics[width=0.32\textwidth]{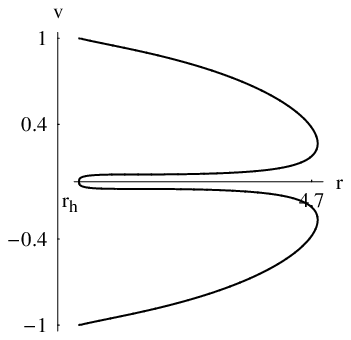}\\
\caption{Accretion of a polytropic test fluid. Contour plots of the Hamiltonian~\eqref{p17} for $C_{5}=X=1$, and $1<\ga =5/3<2$ showing non-global solutions. The solutions cross the $r$ axis at $r=r_h=3^{1/3}$~\eqref{p23}. Left plot corresponds to $Z=9$, $\mathcal{H}=\mathcal{H}_{c1}=53.7813$ and the four CPs are ($r_{c1}=1.92371,v_{c1}=0.715054$), ($r_{c1},-v_{c1}$),  ($r_{c2}=3.27018,v_{c2}=0.602669$), and ($r_{c2},-v_{c2}$). The CPs ($r_{c2},v_{c2}$) and ($r_{c2},-v_{c2}$) are not part of the solution curve $\mathcal{H}=\mathcal{H}_{c1}$, for $\mathcal{H}_{c2}\neq \mathcal{H}_{c1}$. Middle plot corresponds to $Z=Z_0=6.78181083$ for which each couple of CPs of the same sign of $v$ merge with $\mathcal{H}_{c}=35.8097$ and ($r_{c}=2.351,v_{c}=0.6482$). The black, magenta, and blue curves correspond to $\mathcal{H}=\mathcal{H}_{c}$, $\mathcal{H}=\mathcal{H}_c+3$, and $\mathcal{H}=\mathcal{H}_c+10$, respectively. Right plot corresponds to $Z=1$ and $\mathcal{H}=20$ with no CPs.}\label{Fig5}
\end{figure*}

\begin{figure}[!htb]
\centering
\includegraphics[width=0.45\textwidth]{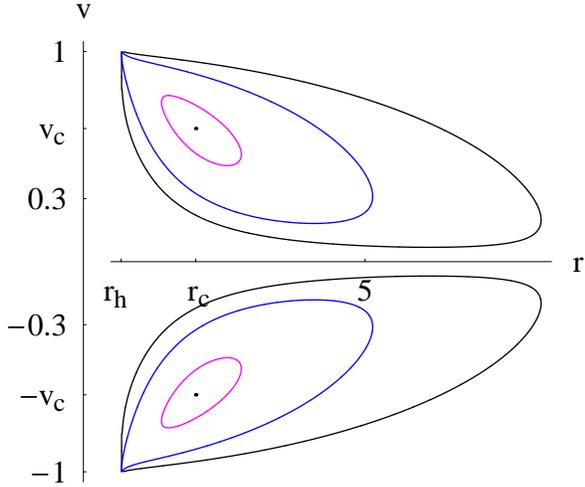}\\
\caption{Accretion of a polytropic test fluid. Contour plots of the Hamiltonian~\eqref{p17} for $C_{5}=X=1$, $Z=9$, and $\ga =7/3>2$. The solution does not cross the $r$ axis. The magenta, blue, and black curves correspond to $\mathcal{H}=\mathcal{H}_c+1$, $\mathcal{H}=\mathcal{H}_c+10$, and $\mathcal{H}=\mathcal{H}_c+30$, respectively, with $\mathcal{H}_c=11.8888$. The plot of $\mathcal{H}=\mathcal{H}_c$ is made of the two CPs ($r_{c}=2.53004,v_{c}=0.633548$) and ($r_{c},-v_{c}$).}\label{Fig6}
\end{figure}

Using~\eqref{b10} or, preferably, the general expression~\eqref{b10b}, in~\eqref{pl3} we arrive at
\begin{equation}\label{pl5}
    h=m\Big[1+Z\Big(\frac{1-v^2}{r^4Av^2}\Big)^{(\ga-1)/2}\Big],
\end{equation}
where
\begin{equation}\label{pl6}
   Z\equiv \frac{K\ga }{m(\ga-1)}\Big|\frac{C_1}{c_3}\Big|^{\ga-1}=\text{ const.}>0
\end{equation}
is a positive constant. If the CPs exist, $Z$ takes the special form
\begin{equation}\label{pl6b}
   Z\equiv \frac{K\ga n_c^{\ga-1}}{m(\ga-1)}~\Big(\frac{r_c^5A_{c,r_c}}{4}\Big)^{(\ga-1)/2}=\text{ const.}>0.
\end{equation}

Inserting~\eqref{pl5} into~\eqref{h3} we evaluate the Hamiltonian by
\begin{equation}\label{p17}
   \mathcal{H}=\frac{A}{1-v^2}~\Big[1+Z\Big(\frac{1-v^2}{r^4Av^2}\Big)^{(\ga-1)/2}\Big]^2,
\end{equation}
where $(c_3m)^2$ has been absorbed into a re-definition of ($\bar{t},\mathcal{H}$).

The constraint $X>0$, in~\eqref{1}, yields $A_{,r}>0$ for all $r$, and this implies that the constant $Z>0$ (recall that $\ga>1$). Thus, the sum of the terms inside the square parentheses in~\eqref{p17} is positive while the coefficient $A/(1-v^2)$ diverges as $r\to\infty$ ($0\leq 1-v^2<1$). So, the Hamiltonian too diverges as $r$ approaches spatial infinity. Since the Hamiltonian has to remain constant on a solution curve, we conclude that there are no global solutions (solutions that extend to, or emanate from, spatial infinity). This conclusion is general and it extends to all anti-de Sitter-like solutions~\cite{Ahmed:2015tyi}.

Since $\ga>1$, the solution curves do not cross the $r$ axis at points where $v=0$ and $r\neq r_h$, for otherwise the Hamiltonian~\eqref{p17} would diverge there. The curves may cross the $r$ axis at $r=r_h$ only. The horizon~\eqref{3a} being a single root to $A(r)=0$, if we assume $v\propto |r-r_h|^{\de}$ and $\de >0$ near the horizon, it is easy to show that
\begin{equation}\label{p23}
    |v|\propto|r-r_h|^{\frac{2-\ga}{2(\ga-1)}},
\end{equation}
that is, $\de =(2-\ga)/[2(\ga-1)]$. Eq.~\eqref{p23} being \emph{valid} for $\de >0$, we see that only physical solutions with $1<\ga<2$ may cross the $r$ axis. For these values of $\ga$, the pressure $p=Kn^{\ga}$ diverges at the horizon as
\begin{equation}\label{p23b}
    p\propto |r-r_h|^{\frac{-\ga}{2(\ga-1)}}\qquad (1<\ga<2).
\end{equation}

Now, substituting
\begin{equation*}
    Y=m(\ga-1)Z\Big(\frac{1-v^2}{r^4Av^2}\Big)^{(\ga-1)/2}
\end{equation*}
into~\eqref{pl3}, we arrive at
\begin{equation}\label{p18}
a^2=Z(\ga -1-a^2)\Big(\frac{1-v^2}{r^4Av^2}\Big)^{(\ga-1)/2},
\end{equation}
which along with Eq.~\eqref{cp2} take the following expressions at the CPs
\begin{align}
\label{p19}&v_c^2=Z(\ga -1-v_c^2)\Big(\frac{1-v_c^2}{r_c^4A_cv_c^2}\Big)^{(\ga-1)/2},\\
\label{p20}&v_c^2=\frac{2Xr_c^3+3C_5}{6Xr_c^3-9C_5},
\end{align}
where we have used~\eqref{1} to reduce the rhs of~\eqref{cp2}. For a given value of the positive constant $Z$, the resolution of this system of equations in ($r_c,v_c$) provides all the CPs, if there are any, the values of which are then used to determine $n_c$ from~\eqref{pl6b}.

Numerical solutions to the system of equations~\eqref{p19} and~\eqref{p20} are shown in Figures~\ref{Fig5} and~\ref{Fig6}. The constant $Z$ is a collection of parameters depending on the black hole and the barotropic fluid. For a given black hole solution, $Z$ is roughly proportional to $Kn_c/m$. For the physical case one is generally interested in astrophysics, $1<\ga<2$, the solution curve has two CPs of the same sign of $v$ for large values of $Z$ (in total four CPs as in the left plot of Fig.~\ref{Fig5}). As $Z$ reaches some critical value, $Z_0$, each couple of CPs of the same sign of $v$ merge as in the middle plot of Fig.~\ref{Fig5}. Below that critical value of $Z$ there are no CPs as in the right plot of Fig.~\ref{Fig5}. For $Z\geq Z_0$, we have heteroclinic flow between two CPs of same value of $r_c$ and opposite values of $v_c$.

The critical flow in the left plot of Fig.~\ref{Fig5} is no difference of that of Fig.~\ref{Fig2} (black plot). The only different feature is that the former flow in non-global while the latter flow is global. Similarly, the magenta and blue curves (corresponding to $\mathcal{H}>\mathcal{H}_c$) of the left and middle plots of Fig.~\ref{Fig5} have branches which are subsonic for the whole process of accretion-flowout as is the case of the curves of Fig.~\ref{Fig2} corresponding to $\mathcal{H}>\mathcal{H}_c$. Another similarity emerges upon comparing the solutions with no CPs corresponding to $Z<Z_0$ (right plot of Fig.~\ref{Fig5}) with those of Fig.~\ref{Fig3} where no CPs occur too.

A common conclusion we can draw upon comparing the solutions of this section with those of the previous one is that low pressure fluids ($k$ and $K$ small) do not develop critical flows (no CPs) and high pressure fluids develop critical flows but they may maintain purely subsonic, even non-relativistic, flows.

Barotropic fluids with $\ga>2$, if there are any, may have CPs but no critical flow and their accretion velocity never vanishes as depicted in Fig.~\ref{Fig6}. The accretion make take place along two different paths starting from rightmost point of the lower branch of Fig.~\ref{Fig6}. For large values of the Hamiltonian (this would be the case if $Z$ is large, $n$, or $K$), the accretion along one of these two paths is almost non-relativistic for $r>r_c$, then the velocity jumps to supersonic and relativistic values as $r$ approaches $r_h$. For lower values of the Hamiltonian, the accretion takes place near the CP and the polytropic fluid never reaches the horizon.

As the title of this section indicates, the analysis made in this section and in the previous ones concern accretion of \emph{test} fluids neglecting all backreaction effects. This rules out any homoclinic flow and motion along closed paths, as those shown in Fig.~\ref{Fig6}, where $v$ conserves the same sign but $r$ increases and decreases.

\section{Conclusions}

In this paper, we discussed in detail the accretion process of  a spherically symmetric black hole in the context of $f(T)$ gravity. In order to select the form of $f(T)$ model, we adopted the Noether Symmetry Approach,  following  \cite{Paliathanasis:2014iva}. In particular, we discussed spherically symmetric  solutions coming from $f(T)=T^m$ models (and, in general,  analytic $f(T)$ models) that give rise to metrics of the form \eqref{3m} and related gravitational potentials of the form \eqref{mfr}, see \cite{Paliathanasis:2014iva} for details.

We have analyzed the motion of isothermal relativistic and ultra-relativistic fluids by means of a Hamiltonian dynamical system capable of representing  hydrodynamics around the black hole.  The thermodynamical properties of the fluids have been discussed according to the suitable EOS. Furthermore conserved quantities and CPs have been selected for any fluid. Roughly, the accretion mechanism can be classified as subsonic and supersonic according to the features of the black hole and the EOS. In particular, the 3-dimensional velocity flow strictly depends on the EOS, the radius and the CPs on the phase space. Finally, the results have been compared to the analogue results in $f(R)$ gravity putting in evidence similarities and differences.

Clearly, the accretion process of the fluids flowing the black holes strictly depends on the conserved quantities (Noether's symmetries) and the structure of CPs, as shown above.  If conserved quantities are not identified, it could become  extremely difficult to  define the phase space structure of the dynamical problem and consequently the features of  CPs. In conclusion, identifying the Noether symmetries allows to fix the model (i.e. the form of $f(T)$), to derive the metric and the gravitational potentials,  thanks to the reduction of the dynamical system, to define the form of the space phase. Models without these features are very difficult to be handled.

From a very  genuine observational point of view, these studies could be related to the possible observable features of $f(T)$ black holes.  In particular, the possibility to investigate $f(T)$ vs $f(R)$ black holes could be a powerful tool to discriminate between the curvature (GR) and torsional (TEGR) formulation of theories of gravity (see \cite{Cai:2015emx} for a detailed discussion). {Specifically, the accretion process onto  a black hole could be the feature capable of discriminating among competing models and, in general, between a curvature or a torsional formulation. A main role in this discussion is played by the stability conditions. For example, as discussed in \cite{mik} for the case of $f(R)$ gravity, the stability conditions for any self gravitating object  strictly depend on the theory. There is demonstrated that the Jeans stability criterion is different if one consider $f(R)$ instead of GR because effective mass, stability radius, Jeans wave length and the other parameters characterizing any astrophysical object slightly change according to the underlying model. In general,  if the accretor has a mass $M$ and a radius $\cal R$, the gravitational energy release is
\begin{equation}
\label{accretion1}
\Delta E_{acc}=\frac{GM}{\cal R}\,.
\end{equation}
Clearly the accretion yield increases with the compacteness $M/{\cal R}$, that is, given a mass $M$, the yields depends on the accretor radius.
Considering alternative theories of gravity, the above relation can be written as
\begin{equation}
\label{accretion2}
\Delta E_{acc}=\frac{G_{eff}M}{\cal R}\,,
\end{equation}
where the features of the given model can be summarized into the effective gravitational coupling $G_{eff}$. This means that the effective potential (related, for example, to the $g_{00}$ component of the metric), determines the accretion process.
For example, the potential \eqref{mfr} indicates that the extra terms with respect to the Newtonian one contribute to any accretion process by modifying the accretion yield. As discussed in Sec.VI D, differences and similarities between the $f(T)$ and $f(R)$ pictures can be put in evidence by a detailed study of the accretion process. In particular, the number of CPs, the state parameter $k$ and other features, besides the effective potential, can discriminate among competing models.
From a genuine observational point of view, luminous phenomena powered by  black holes could contain features capable of discriminating among  theories as soon as the parameters $G$, $M$, and $\cal R$ are combined into a gravitational potential.
For example, the accretion luminosity
\begin{equation}
\label{lum}
 L_{acc}=\frac{GM}{\cal R}\dot{M}=\eta c^2 \dot{M}\,,
\end{equation}
is a feature directly related to these phenomena. Here $\dot{M}$ is the mass variation with time. If one consider a gamma ray burst, we have $L\sim 10^{52}$ $erg/s$ with $\dot{M}\sim 0.1 M_{\odot}/s$. As shown in \cite{gae}, this huge amount of energy can be addressed in a strong field regime by curvature corrections. In other words, the role of $G_{eff}$ for the  adopted underlying model, is crucial. Furthermore, other characterizing parameters, besides $G_{eff}$, can be identified to discriminate observationally concurring accretion models: e.g. the Salpeter timescale \cite{salpeter}, blackbody temperature$T_b$ for thermalization, Eddington limit  \cite{eddington} and so on.
These arguments will be the topic of a forthcoming paper.}

\begin{acknowledgments}
S.B. is supported by the Comisi{\'o}n Nacional de Investigaci{\'o}n Cient{\'{\i}}fica y Tecnol{\'o}gica (Becas Chile Grant No.~72150066).
S.C. is supported by INFN ({\it iniziativa specifica} TEONGRAV) and acknowledges the
COST Action CA15117 (CANTATA).
\end{acknowledgments}

\end{document}